\documentclass[preprint,authoryear,12pt]{elsarticle}
\usepackage[usenames,dvipsnames]{xcolor}

\usepackage{graphicx}

\usepackage{amssymb}

\usepackage[ps2pdf,%
a4paper=true,%
breaklinks=true,%
colorlinks=true,%
pdfauthor={First Author et al.},%
pdftitle={Template for manuscripts in Advances in Space Research}%
]{hyperref}

\journal{Advances in Space Research}

\begin{document}

\begin{frontmatter}



\title{Off-axis irradiation and the polarization of\\broad emission lines in active galactic nuclei}

\author[strasb]{Ren\'e W. Goosmann}
\address[strasb]{Observatoire Astronomique de Strasbourg, Universit\'e de Strasbourg, CNRS, UMR 7550, 11 rue de l'Universit\'e, 67000 Strasbourg, France}
\ead{rene.goosmann@astro.unistra.fr}

\author[valpo,scruz]{C. Martin Gaskell}
\address[valpo]{Centro de Astrof\'isica de Valpara\'iso y Departamento de F\'isica y Astronom\'ia, Universidad de Valpara\'iso, Chile}
\address[scruz]{Department of Astronomy and Astrophysics, University of California, Santa Cruz, CA 95064, USA}

\author[strasb,prag]{Fr\'ed\'eric Marin}
\address[prag]{Astronomical Institute of the Academy of Sciences, Bo{\c c}n\'i II 1401, CZ-14100 Prague 4, Czech Republic}

\begin{abstract}
The {\sc stokes} Monte Carlo radiative transfer code has been extended to model the velocity dependence of the polarization of emission lines. We use {\sc stokes} to present improved modelling of the velocity-dependent polarization of broad emission lines in active galactic nuclei.  We confirm that off-axis continuum emission can produce observed velocity dependencies of both the degree and position angle of polarization.  The characteristic features are a dip in the percentage polarization and an S-shaped swing in the position angle of the polarization across the line profile.  Some differences between our {\sc stokes} results and previous modelling of polarization due to off-axis emission are noted.  In particular we find that the presence of an offset between the maximum in line flux and the dip in the percentage of polarization or the central velocity of the swing in position angle does not necessarily imply that the scattering material is moving radially. Our model is an alternative scenario to the equatorial scattering disk described by \citet{smith2005}. We discuss strategies to discriminate between both interpretations and to constrain their relative contributions to the observed velocity-resolved line and polarization.
\end{abstract}


\end{frontmatter}

\parindent=0.5 cm

\section{Introduction}

As far as we know, all thermal active galactic nuclei, which we will refer to here simply as ``AGNs'', have a broad-line region (BLR) (see \citealt{antonucci2012} for a discussion of thermal versus non-thermal AGNs). The standard model assumes that all AGNs have  BLRs but that in so-called type-2 AGNs the BLR is obscured by optically-thick dust located on our line of sight, while in type-1 AGNs we can see the BLR directly. Hence, the obscuring medium must be anisotropically distributed around the supermassive black hole and its accretion disk. In the simplest version of orientation unification \citep{antonucci1993}, the obscuring medium is assumed to have a toroidal geometry. The BLR and accretion disc are the inner extension of the accretion flow from the torus \citep{gaskell2008a,gaskell2009}.

It has generally been assumed in the unified scheme that the AGN is symmetric with respect to the axis of the circumnuclear dust. While this is a very natural assumption, there is now growing evidence that this is not the case. For example, a detailed spectroscopic analysis of the ionized outflows in the Seyfert-2 galaxy NGC~1068, suggests that the winds are not expelled along the axis of the torus (which is presumably close to the accretion disk axis) but are misaligned with respect to these axes \citep{raban2009}. Furthermore, \citep{gaskell2008b,gaskell2010,gaskell2011} argues that the energy generation of AGNs is fundamentally non-axisymmetric. The standard accretion disk model states that the spectral energy distribution in the optical/UV, the so-called big blue bump, is a continuous superposition of black body spectra following the radial temperature profile of the disk (see, for example Figure~5 of \citealt{gaskell2011}). As a consequence, variability of different spectral regions corresponds to variations at different disk radii. Because information cannot travel faster than light it is impossible for a given annulus of the disk to vary in a coherent manner and to produce the variability at a given wavelength on the observed time scales. Therefore, the energy dissipation must happen on spatial scales significantly smaller than the corresponding radius of the annulus, i.e. {\em the radiation is emitted in a rather localized region}. This idea laid the ground for the off-axis variability model presented in \citet{gaskell2010,gaskell2011}. This new interpretation of continuum and broad line variability solves a range of problems including line profiles and line profile variability \citep{jovanovic2010}, the conflicting kinematic signatures of the red and blue wings of broad lines observed in velocity-resolved reverberation mapping, the puzzling changes in reverberation lags between different continuum events, and the independent variability of different parts of BLR line profiles \citep{gaskell2008b,gaskell2010}.

Polarization is highly sensitive to departure from rotational symmetry about the line of sight in the emission and reprocessing geometry of an astrophysical object and therefore the polarization percentage and position angle provide important information about asymmetries. The motions of the BLR are predominantly coherent. The dominant motion is rotation, with significant turbulence (see \citealt{gaskell2009} for a review of the BLR, probably with inflow \citep{gaskell+goosmann2013}.  As a result of this there is a correspondence between a given velocity in the line profile and a region in the disc (see Figure 6 of \citealt{gaskell2010}).  Because of this, the off-axis irradiation model makes strong predictions for the velocity dependence of the polarization of the broad emission lines.  In this paper, we present some preliminary modeling of the expected intensity and polarization of broad emission lines produced by a BLR that is illuminated by both axially symmetric and off-axis sources. We provide the modeled intensity and polarization profile of the line as a function of the azimuthal position of the off-axis source and compare our results to spectropolarimetric data for type-1 AGN from the literature.  We have been awarded VLT time for making further spectropolarimetric observations of line profiles and we will give a more detailed analysis of these and other data in subsequent work.

The remainder of the paper is organized as follows: in Sect.~\ref{sec:model}, we describe our model setup and the different modeling cases we consider. The results are then given in Sect.~\ref{sec:results} and discussed in the context of spectropolarimetric data in Sect.~\ref{sec:discuss}. We draw some conclusions and give further perspectives of our work in Sect.~\ref{sec:conclude}.

\section{Model setup}
\label{sec:model}

We model the spectropolarimetric signature of a broad emission line as it appears when the BLR is irradiated simultaneously by an axisymmetric continuum source and an off-axis one. The radiative transfer simulations are conducted with a new version of {\sc stokes} \citep{goosmann2007,marin2012}. The new version, which is not yet public, includes the necessary atomic physics to account for photo-absorption and radiative recombination. In our model, the BLR has the geometry of a flared disk centered on the origin of the model space. We define a spherical coordinate system, with the polar angle, $i$, measured from the symmetry axis of the BLR. The inner and outer boundary of the BLR are sphere segments with radii $r_{\rm out} = 4 \times r_{\rm in}$. Their half-opening angle of $\theta_{\rm BLR} = 25^\circ$ is measured with respect to the equatorial plane defined by $\cos{i} = 0$. The continuum source emits a spectrum that is constant in wavelength and the intensity is equally split between two point-like locations. The ``symmetric'' part comes from the origin of the model space, whereas the ``off-axis'' part is emitted by a point source situated inside the equatorial plane and at the inner edge of the BLR. The azimuthal orientation of the off-axis source with respect to the direction of the observer is measured by the angle $\phi$. For $\phi_0 = 0^\circ$, the source lies ``in front of'' the symmetric source, at its closest approach to the observer (inferior conjunction). For $\phi = 180^\circ$ the off-axis source is at superior conjunction with the origin. The polar viewing angle, $i_0$, of the observer is set to $\mu_0 = \cos{i_0} = 0.875$ or $i_0 \sim 29^\circ$. Around $i_0$ and $\phi_0$, photons are integrated over $\Delta \mu = 0.05$ and $\Delta \phi = 15^\circ$.

We assume that photons can be absorbed at any position inside the BLR. At the absorption edge, the radial optical depth to photo-absorption between $r_{\rm in}$ and $r_{\rm out}$ is set to unity. After an absorption event, instantaneous recombination is assumed and a new, unpolarized line photon is emitted at the same location and into an arbitrary direction. Apart from photo-absorption, the BLR also allows for Thomson scattering with a radial optical depth of $\tau_{\rm Th} = 0.24$. All absorbing, re-emitting and scattering material shares a common rotational velocity field.  Following \citet{gaskell2010} this is assuming that the BLR can be modelled as a collection of small clouds on Keplerian orbits at all possible inclinations $-\theta_{\rm BLR} < \theta < \theta_{\rm BLR}$ and orientations of the line of node. The tangential velocity along the orbits scales with $r^{-0.5}$ and is normalized to $1400$~km/s at $r_{\rm in}$. The off-axis source shares the Keplerian rotation of the inner edge of the BLR. For a $10^5$ solar mass black hole, the orbital period at the inner edge of the BLR assumed in this model is about 1~year. Hence, even for a relatively low mass of the supermassive black hole the azimuthal position of the off-axis source remains constant over typical spectropolarimetric observation times. Note, that for the results presented in this paper, no inflow or outflow velocities of the BLR are assumed.

\section{Results and interpretation}

\subsection{Line profile and polarization at different azimuths}
\label{sec:results}

\begin{figure}[t]
  \centering
  \includegraphics[width=\textwidth]{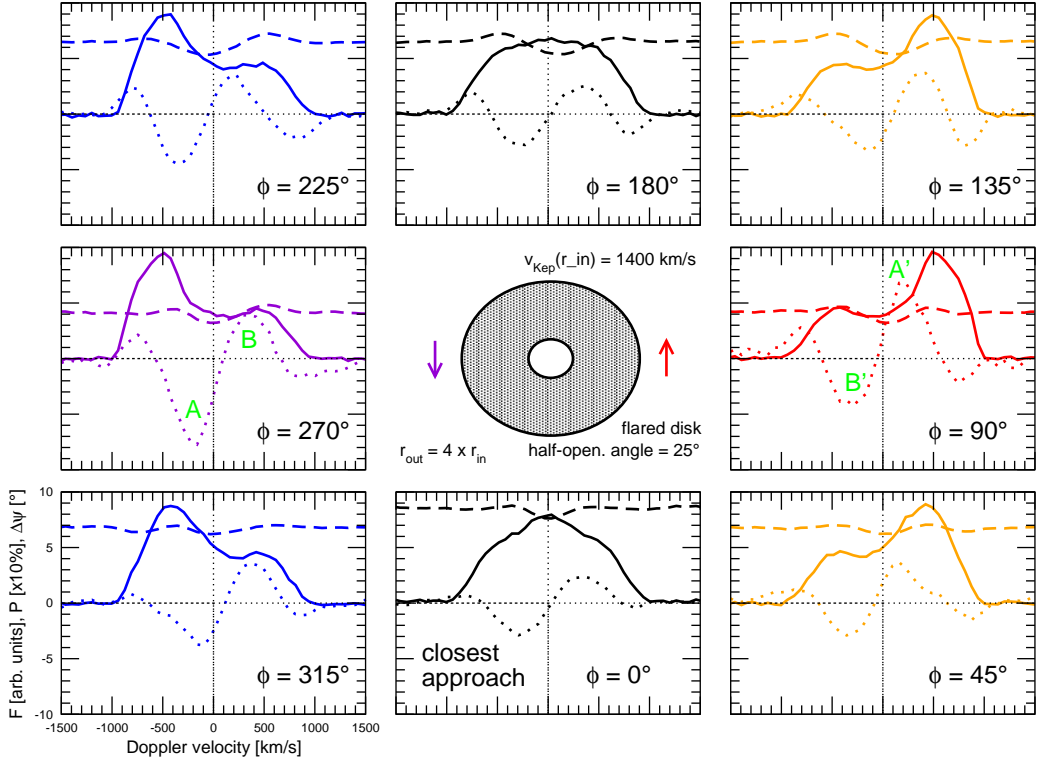}
  \caption{Modeling the spectropolarimetric appearance of a broad emission line as a function of the azimuthal phase $\phi$ of an off-axis continuum source. The eight panels show the profile of the line flux, $F$ (solid line), the polarization percentage, $P$ (dashes), and departure of the polarization position angle, $\Delta \Psi$ (dots) from the overall continuum polarization in Doppler velocity space.}
  \label{fig1}
\end{figure}

Some of our spectropolarimetric modeling results as a function of the azimuthal angle $\phi$ of the off-axis source are shown in Fig.~\ref{fig1}. We plot the total flux spectrum, $F$, the percentage of polarization, $P$, and the deviation of the polarization position angle from its mean, $\Delta \psi$, as a function of observed radial velocity across the line profile. The overall level of polarization $P = 0.7\%$  which is typical of the observed line polarization (see, for example, \citealt{smith2002}). Although it is not shows in the figure, we find that the position angle of the continuum polarization at inferior and superior conjunction is parallel with respect to the projected disk axis. This is as is expected in type-1 objects without scattering in polar winds \citep{goosmann2007,marin2012}. The continuum polarization then varies within $\pm 10^\circ$ from the parallel polarization as a function of the azimuthal position of the off-axis source. This behavior is related to the asymmetric irradiation and will be studied in more detail in future work.

We confirm the polarization behavior predicted in Figure~9 of \citet{gaskell2010}: the percentage of polarization drops across the line centroid and the position angle shows a characteristic S-shaped variation with respect to its mean. The details of this swing in position angle across the line depend on the azimuthal position and are symmetric between the approaching and receding side of the BLR. We mark the corresponding features A, A' and B, B' for phases $\phi = 90^\circ$ and $\phi = 270^\circ$ in the plot.

It can be seen that there is a systematic behavior of the spectral and polarization properties as a function of the phase angle $\phi$. In the line, the spectral flux is dominated by unpolarized radiation from recombination. When the symmetric and the off-axis source are close to the line of sight ($\phi = 0^\circ$ or $\phi = 180^\circ$), the line shape is symmetric and its width represents the integration over all projected radial Doppler velocities apparent in the BLR.  The line profile differs between inferior and superior conjunction because the observer is inclined at $i \sim 29^\circ$ and sees a different irradiation geometry between $\phi = 0^\circ$ and $\phi = 180^\circ$. Towards maximum elongation, at $\phi = 90^\circ$ or $\phi = 270^\circ$, the irradiation of either the approaching or the receding side of the BLR is stronger and the profile becomes significantly asymmetric as is often observed.

\begin{figure}[t]
  \centering
  \includegraphics[width=0.6\textwidth]{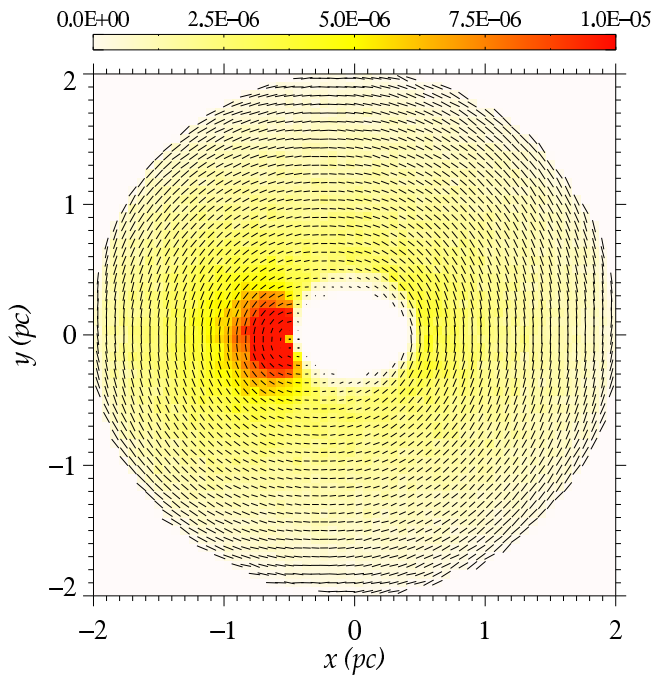}
  \caption{Modeling the polarized continuum flux (color-coded in arbitrary units) as well as polarization percentage and position angle (length and orientation of the black bars) across the accretion disk seen at $i \sim 29^\circ$ for off-axis irradiation at $\phi = 270^\circ$. To obtain this polarization map, {\sc stokes} was run in imaging mode.}
  \label{fig2}
\end{figure}

The wavelength-independent contribution to the polarization is produced by Thomson-scattered continuum radiation and is in agreement with our previous modeling of an equatorial scattering region \citep{goosmann2007,marin2012} as well as with the observed polarization degree in Seyfert-1 galaxies \citep[see e.g.][]{smith2002}. Note that the line polarization is much more sensitive to the orbital phase of the off-axis emission than the continuum emission because different parts of the line profile correspond to localized regions in the disc while the continuum polarization results from integrating the scattered flux over the entire BLR. More polarized flux comes from parts that are closer to the off-axis source and therefore the polarization position angle of the continuum is different at all phases (this is not apparent in Fig.~\ref{fig1} where we only show the deviation of $\psi$ from its phase-dependent mean). The spatial distribution in the resulting polarization position angle across the disk together with the emerging polarized continuum flux is shown in Fig.~\ref{fig2} where {\sc stokes} was applied in imaging mode.

The wavelength-dependent variation of the polarization across the broad line is related to the selective, velocity-dependent dilution by unpolarized line flux occurring in different parts of the BLR since a given velocity comes from specific regions of the BLR moving at the same speed when projected onto the line of sight \citep[see Fig.~6 in][]{gaskell2010}. This dilution of the polarization by unpolarized light can have two effects depending on the amount of unpolarized line flux. In the line wings, the re-emission is still weak and the selective dilution of parts of the BLR leads to a less symmetric distribution in polarized flux and significant net polarization. In the line centroid, however, the dilution due to the unpolarized line photons becomes strong enough to efficiently lower the net polarization despite the larger asymmetry in polarized flux. This mechanism explains why the percentage of polarization rises in the line wings but then diminishes in the centroid. Note that the depth of the polarization dip across the line centroid is much shallower than in the observations. This is because we do not include the unpolarized narrow component of the line in our modeling. The narrow component comes from much more distant and more slowly moving regions of the AGN and thus it cannot alter the polarization properties in the broad line wings. Dilution of the centroid polarization is the only effect to be expected from the narrow lines.

When carefully examining the spectropolarimetric modeling shown in the figure, a slight asymmetry of the polarization at $\phi = 0^\circ$ and $\phi = 180^\circ$ or at $\phi = 90^\circ$ and $\phi = 270^\circ$ becomes apparent. This is related to the Doppler shifting around the absorption edge as well as across the line. The continuum spectrum emitted by the two sources is flat in terms of spectral intensity, $F_{\rm C} \, {\rm d}\lambda \propto \lambda^0 {\rm d}\lambda = \rm const$, and thus the photon index is unity, $N_{\rm ph,C} \, {\rm d}\lambda \propto \lambda^1 {\rm d}\lambda$. On the approaching side of the BLR, the incident continuum as seen in the reference frame of the disk material therefore differs from the one at the receding side. As a consequence, the number of re-emitted line photons is not exactly the same for positive and negative radial velocities. A minor contribution to the asymmetry also comes from the fact that the Doppler shifts are computed by a relativistic formula, which is not symmetric between blue and redshifting. At the inner edge of the BLR, the Keplerian velocity reaches $\sim 1400$~km/s which is about 0.5\% of the speed of light and very mild relativistic beaming effects become apparent.

\subsection{Comparison to previous results of {\sc bl-resp} and observations}
\label{sec:discuss}

It is instructive to compare the modeling obtained with {\sc stokes} with the less sophisticated polarization modelling results from the {\sc bl-resp} shown in Fig.~9 of \citet{gaskell2010}. Both codes reproduce asymmetric line profiles, the bumps in polarization percentage in the line wings and a dip across the line centroid, as well as the S-wave variation of the polarization position angle. However, there are some differences in the details between the two modeling schemes:

\begin{enumerate}

\item The dip in polarization and the switch of the associated rotation of the polarization position angle do not occur at the same wavelengths. In the case of {\sc bl-resp}, the minimum of $P$ is at the wavelength of the line maximum that can be strongly shifted depending on the azimuthal phase of the off-axis source. According to the results from {\sc stokes}, the minimum of $P$ varies less with $\phi$ even at the phases where a strong asymmetry in polarized flux is expected ($\phi = 45^\circ, 135^\circ, 225^\circ, 315^\circ$).\\

\item The S-wave in polarization position angle across the emission line is more complicated in the {\sc stokes} results. There are secondary half-loops to the S-shape that is not as symmetric as in the case of {\sc bl-resp}.

\end{enumerate}

These mismatches are related to the different polarization mechanisms applied in the two codes. The recombination lines in {\sc stokes} are intrinsically unpolarized and the continuum polarization is induced by additional Thomson scattering. In {\sc bl-resp}, the line emission includes intrinsic polarization calculated according to a scattering phase function.  Also, {\sc bl-resp} does not take into account the effects of multiple scattering that start to become apparent at the Thomson optical depths applied here. Still, it is important to note that both codes agree upon the fact that the resulting spectral and polarization features are mostly geometrical.

Comparison of these modeling results to the spectropolarimetric data is not always straightforward, but some remarkable agreement to our modeling can be found in published spectropolarimetry (e.g., \citealt{smith2002}). The ``double-horn'' profile in $P$ appears clearly in Mrk~6 and Mrk~509 and is less symmetric in PG~1700 and E~1821. The dip in $P$ is associated with the swing of the polarization position angle. The position angle swing in PG~1700 has the opposite sense compared to the other three objects. Note that the {\sc stokes} modeling reproduces the offset in radial velocity between the dip in $P$ and the line maximum. Furthermore, the {\sc stokes} results suggest a complex dependence of $\psi$ with wavelength that is apparent in the data. We relate these properties to the complex coupling of dilution effects due to recombination and subsequent Thomson scattering inside the BLR.

In our modelling so far, for simplicity, we have only considered {\em} one off-axis source of illumination in addition to the usual axisymmetric illumination.  In real AGNs there is no reason to expect that in there will be just one off-axis region flaring at a given time, and indeed there is already evidence from velocity-resolved reverberation mapping \citep{sergeev2001} for simultaneous multiple active regions (see Fig.\@ 3 of \citealt{gaskell2011}).  Multiple simultaneous off-axis flaring regions will obviously complicate the interpretation of spectropolarimetry of broad emission lines.

The complex polarization spectrum across broad emission lines in AGN was explained before in an axis-symmetric geometry \citet{smith2004,smith2005}. In this interpretation the characteristic S-wave and the dip in polarization degree across the line centroid are related to the rotation of the emitting BLR and subsequent scattering inside a coplanar,  equatorial scattering ring lying outside the BLR and/or partly being intermixed with it. This scattering region could be inflowing to explain a blueshift of the line in polarized flux. A strength of our model is that the same model explains redshifted and blueshifted polarized flux which symmetric models need to invoke inflow in the blueshifted cases and outflow in the redshifted cases. In future work, we are going to explore how axis-symmetric models can be discriminated from the off-axis interpretation presented here. However, for blueshifted polarized flux, both models stand on fairly realistic grounds and there is no a priori reason to believe that one necessarily excludes the other.

\section{Conclusions and perspectives}
\label{sec:conclude}

We have incorporated in the {\sc stokes} Monte Carlo radiative transfer code the calculation of emission line profiles resulting from scattering and intrinsic emission. From our modelling we have confirmed that the off-axis continuum emission expected in AGNs produces changes in the percentage and position angle of polarization as a function of radial velocity that are similar to observations. We have shown that the presence of an offset between the maximum in line flux and the dip in $P$ or the swing in ($\psi$) does not necessarily imply that the scattering material is moving radially.

Explaining the polarization profile across broad AGN emission lines by off-axis irradiation offers an alternative interpretation to the scattering disc model introduced by \citet{smith2002,smith2005}. While both models are motivated by a realistic scenario, the question arises how to discriminate one model from the other. We expect polarization variability studies to reveal the difference. At the low end of black hole masses and/or for small inner radii of the BLR, the off-axis source should advance on its orbit on time scales of 1~year or less. If spectropolarimetric monitoring reveals changes of the polarization spectra on these time scales they can be compared to the profile of the polarization percentage and position angle at different azimuthal positions of the off-axis source and, eventually, its Keplerian rotation can be reconstructed. Another scenario takes into account the intrinsic time-dependence of localized continuum sources; a given off-axis source may turn off while a new source at a different azimuthal and/or radial position turns on. Again, polarization variability is the consequence. We believe that such variability effects would be very hard to explain in the framework of an axis-symmetric model.

Of course, it is possible that both, off-axis irradiation and axis-symmetric scattering in an equatorial disc play a role in determining the line polarization. Note that with the axis-symmetric component of the continuum emission and the Thomson scattering inside the flared BLR disc, we adopt here some elements from the model by \citet{smith2005}.  An important future task is to constrain the fraction of continuum that is emitted by one or several (!) bright off-axis sources. In our preliminary modeling presented here we still assume equipartition between both components. In future work we are going to vary their relative contribution and study the effects on the resulting line profiles and polarization. We will give a more detailed discussion of our {\sc stokes} modelling and analysis of specific AGNs in future papers.

\subsection*{Acknowledgements}
We thank the referee, Makoto Kishimoto, for his helpful comments on the manuscript. Furthermore, we are grateful to Luka Popovi\'{c}, Dragana Ili\'c, and Alla Shapovalova for very helpful discussions on the spectropolarimetric profiles of AGN emission lines. This research was supported by the French grant ANR-11-JS56-013-01 of the project POLIOPTIX. MG acknowledges support of this research by the GEMINI-CONICYT Fund of Chile through project 32070017 and by CONICYT of Chile through FONDECYT project 1120957.


\end{document}